# High-Performance Hybrid Electronic Devices from Layered PtSe$_2$ Films Grown at Low Temperature

Chanyoung Yim,[†,‡,#] Kangho Lee,[†,#] Niall McEvoy,[*,†,§,#] Maria O'Brien,[†,§] Sarah Riazimehr,[‡] Nina C. Berner,[†] Conor P. Cullen,[†,§] Jani Kotakoski,[Δ] Jannik C. Meyer,[Δ] Max C. Lemme,[‡] and Georg S. Duesberg[*,†,§]

[†]Centre for Research on Adaptive Nanostructures and Nanodevices (CRANN) and Advanced Materials and BioEngineering Research (AMBER), Trinity College Dublin, Dublin 2, Ireland

[‡]Department of Electrical Engineering and Computer Science, University of Siegen, Hölderlinstraße 3, 57076 Siegen, Germany

[§]School of Chemistry, Trinity College Dublin, Dublin 2, Ireland

[Δ]Faculty of Physics, University of Vienna, Boltzmanngasse 5, A-1090 Vienna, Austria

**ABSTRACT**

Layered two-dimensional (2D) materials display great potential for a range of applications, particularly in electronics. We report the large-scale synthesis of thin films of platinum diselenide ($PtSe_2$), a thus far scarcely investigated transition metal dichalcogenide. Importantly, the synthesis by thermal assisted conversion is performed at 400 °C, representing a breakthrough for the direct integration of this material with silicon (Si) technology. Besides the thorough characterization of this 2D material, we demonstrate its promise for applications in high-performance gas sensing with extremely short response and recovery times observed due to the 2D nature of the films. Furthermore, we realized vertically-stacked heterostructures of $PtSe_2$ on Si which act as both photodiodes and photovoltaic cells. Thus this study establishes $PtSe_2$ as a potential candidate for next-generation sensors and (opto-)electronic devices, using fabrication protocols compatible with established Si technologies.

In recent years, two-dimensional (2D) layered materials, such as graphene and transition metal dichalcogenides (TMDs), have been heavily studied due to their high potential for use in a wide range of future nanoelectronic devices.[1–7] TMDs have the general formula of $MX_2$, where M denotes a transition metal and X is a chalcogen, with weak van der Waals bonding between layers and strong in-plane covalent bonding. Most initial studies on the properties of TMDs used mechanically-exfoliated single- or multi-layered sheets with an inherent lack of scalability.[8–10] However, synthesis of reliable, large-scale TMD thin films is considered a prerequisite for future electronic and photonic device applications. Many studies have shown the possibility of up-scaling production through liquid-phase exfoliation,[11] however the flakes derived from this method typically have small lateral dimensions and a distribution of layer thicknesses.

Recently, continuous, large-scale TMD films, namely of the sulfides and selenides of molybdenum (Mo) and tungsten (W), have been produced by growth techniques based on chemical vapor deposition (CVD) whereby large-area TMD films can be directly synthesized on insulating substrates *via* the reaction of vaporized solid precursors or by thermally assisted conversion (TAC) of pre-deposited metal layers on insulators.[12–14] TMDs synthesized in this manner have shown notable performance in multiple device applications such as chemical/gas sensors and photodetector devices.[15–18] Thus TMDs may provide added functionality for highly-integrated silicon (Si) chips, *e.g.* for applications in the Internet of Things (IoT). To achieve this, compatibility with so-called back-end-of-line (BEOL) processing with temperatures below 450 °C is needed.[19–21] However, typical synthesis temperatures for Mo or W based TMDs exceed 450 °C, even in plasma assisted processes.[22]

There are many other members of the TMD family, including the noble-metal TMDs, which have thus far garnered less attention. Studies have theoretically predicted the properties of these materials and touted them for use in assorted applications,[23,24] but little work has been done to date on their synthesis. Platinum diselenide ($PtSe_2$) is one such material, which has not previously been assessed for use in device applications. Bulk $PtSe_2$ is a semi-metal with zero bandgap,[25,26] which has been synthesized by chemical methods and used as a photocatalytic material in nanocomposites with graphene.[27] Theoretical studies have suggested the emergence of a bandgap in monolayer $PtSe_2$[23,24,28] and a recent experimental study, which outlined the growth of $PtSe_2$ by direct selenization of Pt single crystals under ultra-high-vacuum conditions, verified this and demonstrated that monolayer $PtSe_2$ has a bandgap of 1.2 eV.[26] This thickness-dependent semimetal-to-semiconductor transition suggests that $PtSe_2$ could be a potential candidate for (opto-)electronic device applications.

In this work, we demonstrate the potential of $PtSe_2$ for future hybrid electronic devices, fabricating a high-performance gas sensor and photodetector, as well as a functional photovoltaic cell. Large-scale $PtSe_2$ thin films with controlled thicknesses were directly grown on insulating substrates by TAC. Importantly, the growth temperature is 400 °C, which allows integration of $PtSe_2$ with purpose-designed Si chips. In this manner, a $PtSe_2$ channel was defined and subsequently contacted to fabricate a gas sensor with a chemiresistor structure. This gas sensor showed ultra-high sensitivity to $NO_2$ gas with extremely fast response times. Additionally, large-scale heterostructure diodes have been realized by transferring $PtSe_2$ thin films onto pre-patterned n-type Si (n-Si) substrates at room temperature, forming $PtSe_2$/n-Si Schottky barrier diodes (SBDs). The electrical properties of the devices have been studied through dc current-voltage measurements. Moreover, the photoresponse of the SBDs, including photoconductivity and photovoltaic effects, was carefully examined.

**RESULTS AND DISCUSSION**

*Synthesis and Characterization*

PtSe$_2$ thin films with various thicknesses were prepared by TAC. A growth temperature of 400 °C was used and growth was achieved directly on the Si/SiO$_2$ substrate (in contrast to an earlier work[26]) which makes this process compatible with standard semiconductor back-end-of-line (BEOL) processing. PtSe$_2$ samples with different thicknesses were obtained by modifying the pre-deposited Pt thickness. A photograph of the synthesized PtSe$_2$ films is presented in Figure 1(a).

The atomic structure of the PtSe$_2$ thin film was examined by scanning transmission electron microscopy (STEM) analysis after transferring the films onto TEM grids. A representative high angle angular dark field STEM (HAADF-STEM) image is shown in Figure 1(c) of a PtSe$_2$ thin film synthesized from 0.5 nm Pt starting thickness. This shows the synthesized material to be polycrystalline, with several nanometer sized crystalline domains of varying thickness, similar to what has previously been observed for MoS$_2$.[29] It is possible to identify the atomic structure of the film from the image shown. The brighter spots (Pt atoms) and darker spots (Se atoms) form a 1T crystal structure, similar to that previously observed in layered materials such as HfS$_2$ and CdI$_2$, and in agreement with previous predictions of the crystal structure of PtSe$_2$.[24] A schematic of the 1T crystal structure is shown in Figure 1(b).

Raman spectra of PtSe$_2$ films of varying thicknesses are shown in Figure 1(d). These PtSe$_2$ films were synthesized from Pt films of starting thickness 1, 2, 4 and 5 nm. Two prominent peaks can be identified at ~176 cm$^{-1}$ and 210 cm$^{-1}$, which correspond to the $E_g$ and $A_{1g}$ Raman active modes, respectively. The $E_g$ mode is an in-plane vibrational mode of Se atoms moving away from each other within the layer, while the $A_{1g}$ mode is an out-of-plane vibration of Se atoms in opposing directions. The $E_g$ mode shows a minor blue shift with decreasing film thickness, in agreement with previous work and theoretical calculations.[28] The $A_{1g}$ mode shows a significant increase in intensity with increasing film thickness, due to enhanced out-of-plane interactions from increased layer numbers. The peak at ~230 cm$^{-1}$ can be attributed to an LO mode, similar to those observed in HfS$_2$, ZrS$_2$ and CdI$_2$.[30,31]

X-ray photoelectron spectroscopy (XPS) spectra of the Pt 4f and Se 3d region, acquired on a PtSe$_2$ sample produced by selenizing a 5 nm thick Pt film, are shown in Figure 1(e) and (f), respectively. The Pt 4f is deconvoluted into three contributions, the primary one at ~72.3 eV is attributed to PtSe$_2$ whereas the two smaller ones at ~74 eV and ~71 eV are attributed to oxides and unreacted Pt metal, respectively. The relative atomic percentages of the three identified Pt species indicate that the majority of the Pt has been transformed into the selenide. The Pt 5p$_{3/2}$ peak also lies in the same region, but has not been used for analysis. The Se 3d peak is deconvoluted into two contributions, one from PtSe$_2$ and the other relatively small one at higher binding energies likely stemming from edge or amorphous Se. Additional Raman and XPS spectra acquired from films selenized at different temperatures are presented in the Supporting Information, Figure S.2 and S.3.

*Gas Sensors*

As-grown PtSe$_2$ was used to realize a chemical sensor, or chemiresistor, with ultra-fast response time and sensitivity towards environmental gases at room temperature. It is important to note that the PtSe$_2$ gas sensor is fabricated at 400 °C, which is compatible with BEOL processing, and simple lithography processes (see Methods), making it possible to implement such sensors on top of integrated circuits. As a chemiresistor, the PtSe$_2$ channel had a resistance of approximately 18 kΩ with Ni/Au contact electrodes. The output characteristics ($I_{ds}$ vs. $V_{ds}$ curves) are very linear and the contact resistance derived from 4-probe measurements is small enough to exclude the contribution of contact resistance. In addition, electrical transport measurements of the PtSe$_2$ channel, conducted under back-side gate biases, revealed that the PtSe$_2$ channel showed p-type conduction. (See Figure S.4 (a) and (b) of the Supporting Information.)

Figure 2(a) shows typical gas sensor responses upon periodic NO$_2$ gas exposure, from 0.1 to 1 ppm (parts-per-million) with a bias voltage of 1 V. The electron-withdrawing character of NO$_2$ means that it effectively p-dopes the channel. Note that doping is not used in the conventional terminology as replacing crystal lattice atoms, but rather as electrostatic doping: Adsorbed NO$_2$ molecules on the surface of PtSe$_2$ channel modulate the charge carrier density *via* so-called vicinity doping or adjacent doping. The Fermi level of PtSe$_2$ is shifted toward the valence band, resulting in a resistance decrease.

The PtSe$_2$ based sensor displays ultra-fast response/recovery speed at room temperature compared to state-of-the-art sensors which is important for many applications. As shown in Fig 2(a), it immediately responds upon 10 seconds exposure to 100 sccm flow of NO$_2$ mixture with N$_2$ carrier gas. It is important to note that the actual gas exposure time is much shorter than the applied time due to the limited switching speed of mass-flow controllers, which delays gas exchange in our custom-made sensor test chamber for a few seconds. (See Figure S.5 of the Supporting Information for more details.) Moreover, the original resistance is fully recovered in pure N$_2$ flow at room temperature, without typical post-processes for accelerated recovery speed, such as ultra-violet illumination or annealing. The gas sensor response/recovery characteristics were analyzed by the Langmuir isotherm model,[32,33] $R(t) = R_\infty(1 - e^{-t/\tau})$, where *R(t)* is the resistance at time t, $R_\infty$ is the final resistance produced by equilibrium coverage at approximately 150 Torr, and $\tau$ is the transient response/recovery time. We found that a double exponential model fitted our data well: $R(t) = R_{\infty 1}(1 - e^{-t/\tau_1}) + R_{\infty 2}(1 - e^{-t/\tau_2})$, where $R_\infty = R_{\infty 1} + R_{\infty 2}$ and $\tau = \{R_\infty/(R_{\infty 1}/\tau_1 + R_{\infty 2}/\tau_2)\}$. Each parameter *versus* $NO_2$ concentration upon 10 seconds exposure is depicted in Figure 2(b) and (c). Higher gas concentrations, at a constant exposure time, result in a larger resistance change and overlapping reaction and recovery plots (Figure 2(b)) indicate that the sensors repeatedly fully recover in pure $N_2$ gas. Transient time constants make it possible to quantitate how quickly sensors respond upon gas exposure and recovery. The estimated response and recovery time are 2.0 – 53.7 and 7.4 – 38.7 seconds at 0.1 – 1.0 ppm of $NO_2$ exposure, respectively, which is considerably smaller than previously reported studies of graphene[33–35] and metal-oxide sensors.[36]

The measurement configuration allows $NO_2$ mixtures down to 100 ppb (parts-per-billion). However, signal processing based on the signal-to-noise ratio (SNR) implies even lower limits of detection (LOD). The initial resistance is calculated from the first 100 data points just before the first gas injection and root-mean-square (RMS) noise is derived from the baseline of initial resistance, $RMS_{Noise} = \sqrt{\sum (R - R_0)^2/N}$, where $R$ is measured resistance depending on time, $R_0$ is the initial resistance, and $N$ is the number of data points. According to the linearity of SNR-to-concentration, the theoretical LOD can be extrapolated as SNR must be at least three or larger.[37] With this we estimate that the $PtSe_2$ sensor can reach a level of detection of just a few ppb with less than 10 seconds exposure time. This is a significant improvement on previous gas sensors based on other 2D TMD materials such as $MoS_2$ which often have long response times, up to hours, for low detection limits. For example, M. Donarelli *et al.*[38] presented $MoS_2$-based gas sensors with 20 ppb detection upon 2-hour exposure at 200 °C, but these showed less sensitivity at room temperature. B. Liu *et al.*[39] demonstrated MoS$_2$-based gas sensors with a 20 ppb limit of detection with a shorter exposure time of approximately 15 minutes; however this is still considerably longer than our case.

*Schottky Barrier Diodes*

Schottky barrier diodes were fabricated by transferring PtSe$_2$ films onto pre-patterned n-Si substrates. Compared with the gas sensor device, thicker PtSe$_2$ films (grown from thicker Pt starting layers) were used which formed Schottky contacts with the n-Si substrate (see Methods for details). Typical dc current density-voltage (*J-V*) measurements in the dark of the PtSe$_2$/n-Si SBD device, with a PtSe$_2$ film synthesized by the selenization of a 4 nm thick Pt film, are plotted in Figure 3(c) and (d), showing clear rectifying characteristics. The current transport mechanism of the SBD can be explained by thermionic emission theory[40] and the diode parameters such as the ideality factor (*n*), Schottky barrier height ($\varphi_B$) and series resistance ($R_S$) have been extracted by adopting Cheung's approach.[41] For this device, they were determined to be *n* = 1.9, $\varphi_B$ = 0.71 eV and $R_S$ = 836 Ω. Details of the diode parameter extraction can be found in the Supporting Information. Three more devices with different PtSe$_2$ thicknesses were measured and their diode parameter values were also extracted in a similar manner and summarized in Table 1. These all show low values of ideality factor (1.4 − 1.9), indicating their device performance is good even without optimization. In addition, they all show similar values of Schottky barrier height (~0.7 eV), implying that the interface condition between PtSe$_2$ and Si is nearly identical for all devices. The SBD with a thicker PtSe$_2$ film has a lower value of series resistance which can be attributed to the increase in the volume of the conductive PtSe$_2$ layer.

In the dark and under reverse bias, the PtSe$_2$/n-Si SBDs are in the off-state with low current. Upon exposure to light, a rapid increase of the current density of the SBDs was observed in the reverse bias region. The photoresponse of the devices was measured under illumination of a white-light source onto the surface of the PtSe$_2$ thin film, which the light partially penetrates to reach the interface between PtSe$_2$ and Si. *J-V* characteristics of the PtSe$_2$/n-Si SBD with a PtSe$_2$ film synthesized from 4 nm Pt starting thickness are plotted in Figure 4(a), showing a clear photoresponse in the reverse bias region. Three other SBDs with different PtSe$_2$ thicknesses, which were selenized from 1, 2 and 5 nm thick Pt films, also showed similar photosensitive characteristics and their *J-V* plots are depicted in Figure S.7 of the Supporting Information. The responsivity of the devices was measured at a dc bias of -2 V using a calibrated detector and the maximum responsivity was found to be 490 mA/W at a wavelength of 920 nm. This is comparable to recent results of the graphene/Si heterojunction photodetectors (270 – 730 mA/W),[42–44] and higher than that of multilayer MoS$_2$ photodetectors (10 – 210 mA/W).[29,45,46] The generated photocurrent of the SBD can be attributed to photon-induced mobile charge carriers. When the incident photon energy (*hv*) is greater than the Schottky barrier height and less than the bandgap of the Si ($\varphi_B$ < *hv* < $E_g$, $E_g$ = ~1.1 eV), electrons excited from the PtSe$_2$ can overcome the Schottky barrier and they are injected into the Si. When *hv* is greater than the bandgap of Si ($E_g$ ≤ *hv*), electron-hole pairs are also generated in the depletion layer and Si. In addition, the effect of varying the intensity of illuminating light incident on the SBD was investigated (Figure 4(b)). It was found that with increasing incident light intensity (P$_{in}$) from 0.1 to 16.8 mW/cm$^2$ the photocurrent increases from 0.2 to 3.6 mA/cm$^2$ at a dc bias of -2 V due to the increase of the photon-induced mobile charge carriers. A schematic of the energy band diagram at the junction of the SBD is presented in Figure 4(c). The photocurrent of the four devices with different PtSe$_2$ thicknesses (1, 2, 4 and 5 nm Pt starting thickness) were further investigated under the same light intensity (P$_{in}$ = 12.3 mW/cm$^2$). The photocurrent increases clearly in the reverse bias region as the PtSe$_2$ thickness is increased. The photocurrent in all devices depends on the reverse dc bias (Figure 4(d)), as the increased electrical potential difference across the depletion layer at the junction, due to the external bias, causes stronger acceleration of electrons and holes in the depletion region. Here, it is also evident that devices with thicker PtSe$_2$ films generate higher photocurrents. This indicates that a substantial part of the carrier generation takes place in the PtSe$_2$ films, not the Si substrate. This is in contrast to previously reported SBDs formed between semi-metallic graphene and Si.[47–49] Unlike graphene/Si SBDs, where the graphene is a monolayer thick by definition, the absorbance of the photodiodes presented here can be easily tuned by modifying the initial Pt deposition thickness.

In addition, the PtSe$_2$/n-Si SBDs exhibit a photovoltaic effect. The output characteristics of the open circuit voltage (V$_{OC}$) and short circuit current (J$_{SC}$) for the PtSe$_2$/n-Si SBDs, when illuminated at P$_{in}$ = 4.7 mW/cm$^2$, is depicted in Figure 4(e). Similar to the photoconductivity, thicker PtSe$_2$ films result in higher values of V$_{OC}$ and J$_{SC}$. This can again be attributed to the larger volume of photon absorption at the junction of the device with a thicker film that effectively contributes to the generation of photon-induced mobile charge carriers at the junction. The maximum values of V$_{OC}$ and J$_{SC}$ were found to be 0.17 V and 0.93 mA/cm$^2$, respectively, for the device with the PtSe$_2$ film synthesized from 5 nm Pt. The corresponding maximum output power of the system (P$_{max}$) is ~48.7 µW/cm$^2$ at a voltage of 0.1 V with a current of 0.49 mA/cm$^2$ (Figure 4(f)). Based on this, the fill factor (FF) and power conversion efficiency (PCE) of the device can be extracted. We obtain values of FF = 0.31 and PCE = 1.0 %. The PCE is in fact fairly high for this kind of 2D material class, and it is comparable to previously reported 2D material based Schottky-junction solar cells, which have PCE values of 0.8 – 1.5 % for monolayer-graphene-based SBDs[50,51] and 0.7 – 1.8 % for multilayer-MoS$_2$-based SBDs.[52] While previously-reported graphene-based devices have shown higher PCE values than our PtSe$_2$/n-Si SBD, PtSe$_2$ is grown at a much lower temperature than that typically required for monolayer graphene (~1000 °C). Additionally, one can envisage fabricating PtSe$_2$ channels of controlled thickness directly on SiO$_2$/Si substrates, something which would be highly problematic with graphene. Thus there may be applications where the use of PtSe$_2$ offers significant advantages over graphene, despite its lower PCE. The extracted parameters of the photovoltaic effect for the SBDs are summarized in Table 1. It is known that undesirable parasitics, such as the series resistance ($R_S$) typically associated with real devices, are closely related to the degradation of the PCE in photovoltaic devices.[53] $R_S$ has not been optimized in this first demonstration and arises from the PtSe$_2$ film and the Si substrate, the contact resistances of the metal to PtSe$_2$, which is a well-known bottleneck in 2D materials, or possibly from the presence of an interface layer at the Schottky junction. Our SBDs have $R_S$ values of ~450 to ~1520 Ω, and the devices with higher $R_S$ show lower PCE. Therefore, it is expected that the PCE of the SBDs can be improved by further optimization of the device fabrication process to minimize the effect of $R_S$.

**CONCLUSION**

Large-scale PtSe$_2$ thin films with different thicknesses were synthesized in a controllable manner by thermally assisted conversion of platinum at very low temperature. A gas sensor, with a PtSe$_2$ channel, fabricated using a facile and scalable process, showed ultra-high sensitivity to NO$_2$ with extremely fast response times. In this study, 100 ppb of NO$_2$ was detectable at room temperature and theoretical analysis of the data promises an even lower limit of detection. In addition, Schottky diodes with a vertically-stacked heterostructure were realized by transferring PtSe$_2$ thin films onto pre-patterned Si substrates. These diodes can be utilized as photodetectors as well as photovoltaic cells. The performance of the PtSe$_2$-based device demonstrated in this study is equal or superior to comparable graphene or TMD-based devices. Yet, the low-temperature growth technology introduced here is compatible with current standard semiconductor BEOL processing. Therefore, PtSe$_2$ could be considered as one of the most promising 2D materials for next-generation nanoelectronic device applications.

**METHODS**

*Materials Synthesis*

PtSe$_2$ thin films were synthesized using a TAC process similar to that previously described for MoS$_2$.[14] Pt layers of different thicknesses were sputter coated onto SiO$_2$/Si substrates using a Gatan Precision Etching and Coating System (PECS). The Pt samples were selenized in a quartz tube furnace with two independently-controlled heating zones. Pt samples were loaded in the primary heating zone and heated to 400 °C. The Se source was loaded in the secondary heating zone, which was heated up to the melting point of Se (~220 °C). Ar/H$_2$ (9:1), with a flow rate of 150 sccm, was used to transport the vaporized Se to the Pt samples. A rotary vane pump was used to evacuate the system and keep it under vacuum and the pressure during selenization was typically ~0.7 Torr. A dwell time of 2 hours was used to ensure complete selenization. A schematic diagram of the film synthesis process is presented in Figure S.1 of the Supporting Information.

*Device Fabrication*

The PtSe$_2$ gas sensor was fabricated by simple two-step lithography using a shadow mask. Firstly, a 0.5 nm thick Pt film was sputtered onto SiO$_2$/Si substrates (300 nm thick SiO$_2$) through a shadow mask and selenized by TAC at 400 °C. Secondly, 12 metal electrodes (nickel/gold (Ni/Au) = 20/80 nm) were deposited onto the PtSe$_2$ layer using a shadow mask, which defined a channel with a size of 1 × 0.2 mm$^2$ as shown in the inset of Figure 2(a).

A lightly-doped n-type Si (n-Si) wafer with a thermally-grown silicon dioxide (SiO$_2$) layer (thickness of 150 nm) on top was used as the SBD substrate. The n-Si wafer had a dopant (phosphorus) concentration of 5 × 10$^{14}$ cm$^{-3}$ and <100> orientation. In order to prepare a pre-patterned device substrate, part of the SiO$_2$ layer was etched by 3 % diluted hydrofluoric acid (HF), followed by rinsing with deionized (DI) water. Then, titanium (Ti) and gold (Au) were deposited (Ti/Au = 20/50 nm) on the exposed n-Si and the remaining SiO$_2$ area using a shadow mask to define metal electrodes on the device, forming good ohmic contacts between n-Si and the metal electrodes. The synthesized PtSe$_2$ thin films were transferred onto desired substrates and TEM grids using a typical polymer support transfer technique. Polymethyl methacrylate (PMMA, MicroChem) was spin-coated onto the as-grown PtSe$_2$. The SiO$_2$ layer under the PtSe$_2$ was removed by a wet-etching process using 2 molar sodium hydroxide (2M NaOH) at room temperature. After cleaning in DI water, the PtSe$_2$ with PMMA layers were transferred onto the substrates. The PMMA was removed by immersion in acetone at room temperature for 20 minutes. The native oxide layer on the Si surface was removed using a HF wet-etching process before transferring PtSe$_2$ onto the pre-patterned Si substrates. The SBD has an active area of ~25 mm$^2$ and a schematic diagram and photograph of the fabricated device are shown in Figure 3(a) and (b).

*Characterization*

Raman analysis was carried out using a Witec Alpha 300 R confocal Raman microscope with an excitation wavelength of 532 nm at a power of < 300 µW and a spectral grating with 1800 lines/mm. The spectra shown for each sample were obtained by averaging 10 discrete point spectra. XPS spectra of the Pt 4f and Se 3d core-levels were recorded under ultra-high-vacuum conditions (<$10^{-8}$ mbar) on a VG Scientific ESCAlab MkII system using Al $K_\alpha$ X-rays and an analyzer pass energy of 20 eV. After subtraction of a Shirley background, the core-level spectra were fitted with Gaussian-Lorentzian and Doniach-Sunjic (for the metallic Pt 4f component) line shapes using the software CasaXPS. STEM samples were prepared by transferring the as-grown films onto Quantifoil TEM grids. The STEM studies were performed in a Nion UltraSTEM 100 operated at 60 kV using a HAADF detector. Prior to STEM measurements, the samples were annealed in vacuum for more than 14 hours at 150 °C.

The $PtSe_2$ gas sensor was tested using a custom-made vacuum chamber with remote-controllable mass-flow controllers (MFCs). All the sensors were loaded in the chamber at a constant pressure of approximately 150 Torr with a constant flow of 100 sccm of a gas mixture. 10 ppm of $NO_2$ was introduced and diluted with dry nitrogen through MFCs. The change in resistance upon periodic gas exposure was monitored using a Keithley 2612A source meter unit.

Electrical measurements of the SBDs were performed under ambient conditions using a Karl Suss probe station connected to a Keithley 2612A source meter unit. The metal electrode connected to the $PtSe_2$ on the $SiO_2$ part of the device was positively biased and the electrode on the n-Si was negatively biased. A white light source with a solid state dimmer for variable light intensity (ACE Light Source, SCHOTT: A20500, 150 watt halogen lamp) was used for photoresponse measurements.

## FIGURES

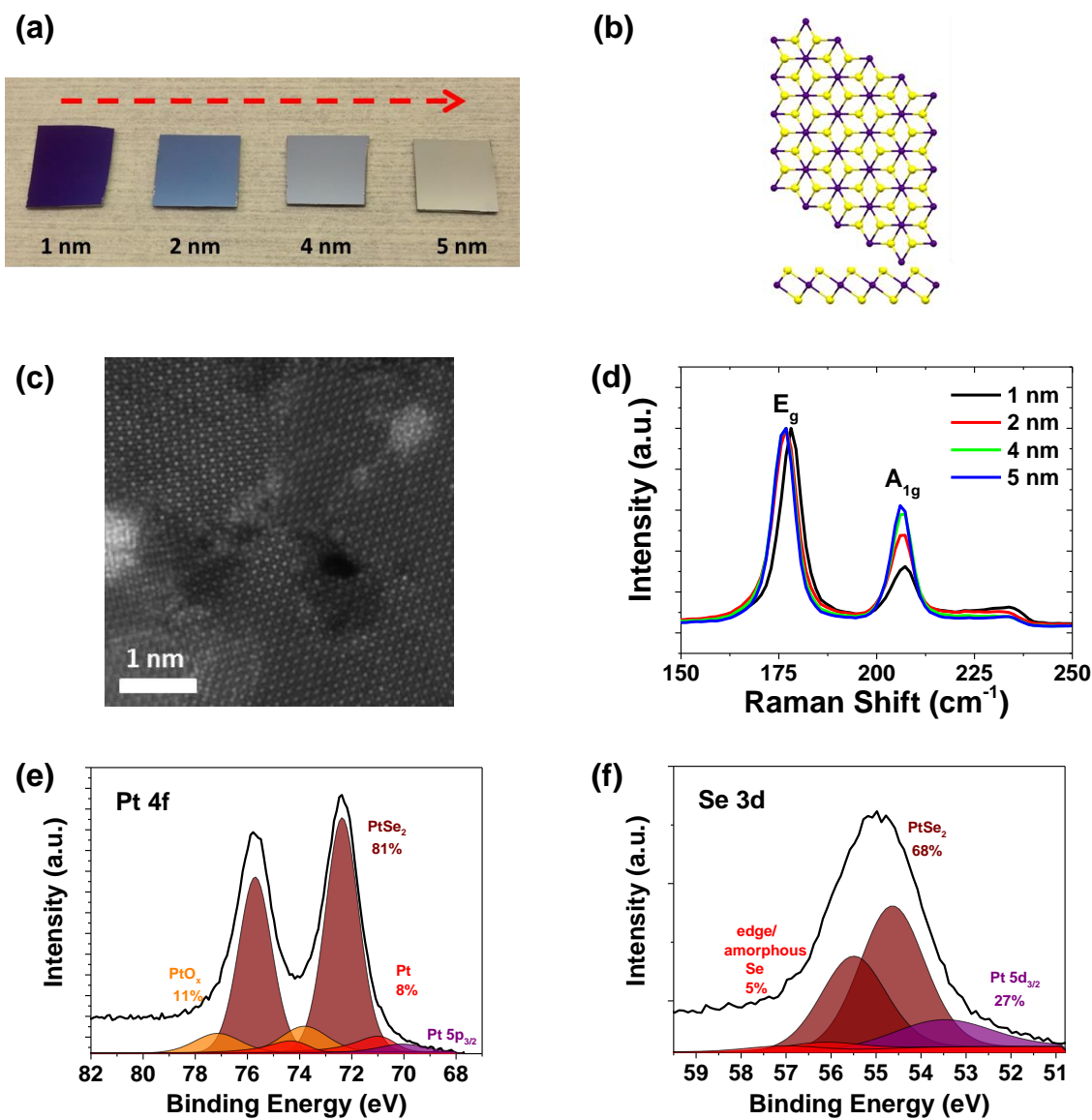

Figure 1. (a) Photograph of the PtSe$_2$ samples with different thicknesses grown on SiO$_2$(300 nm thick)/Si substrates. Initial Pt deposition thicknesses are 1, 2, 4 and 5 nm. (b) Schematic diagram of the 1T crystal structure of PtSe$_2$ (Pt atoms: purple, Se atoms: yellow). (c) HAADF-STEM image of a PtSe$_2$ film synthesized from 0.5 nm Pt starting thickness, transferred onto a holey

carbon grid, showing the 1T structure. (d) Raman spectra of PtSe$_2$ films of different thickness. XPS spectrum for (e) Pt 4f region and (f) Se 3d region of a PtSe$_2$ film produced by selenizing a 5 nm thick Pt film.

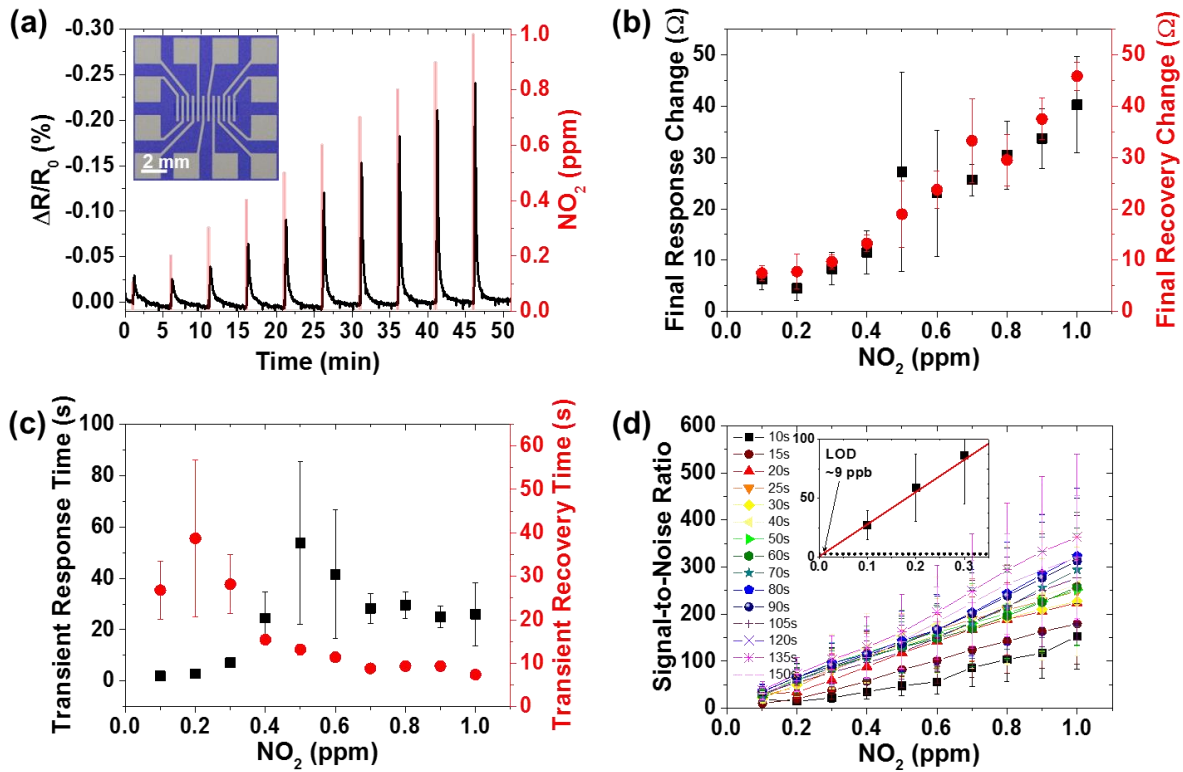

Figure 2. (a) A sensor response plot of percentile resistance change *versus* time of a PtSe$_2$ film produced by selenizing a 0.5 nm thick Pt film at 1 V bias voltage. Red vertical bars indicate NO$_2$ gas injections and black the resistance change of PtSe$_2$ channel. (Inset: Optical micrograph of a contacted sensor device.) (b) The estimated final resistance change and (c) transient response/recovery time are depicted at various NO$_2$ concentrations for 10 seconds exposure. (d) Signal-to-noise ratios are linearly proportional to NO$_2$ concentration in the range of 0.1 to 1 ppm

for 10 to 150 seconds exposure. Inset: An extrapolation of the theoretical limit of detection. Red line is linearly fitted and a broken line indicates a signal-to-noise ratio of 3. All error bars represent the standard deviation of measurements from 8 sensor devices.

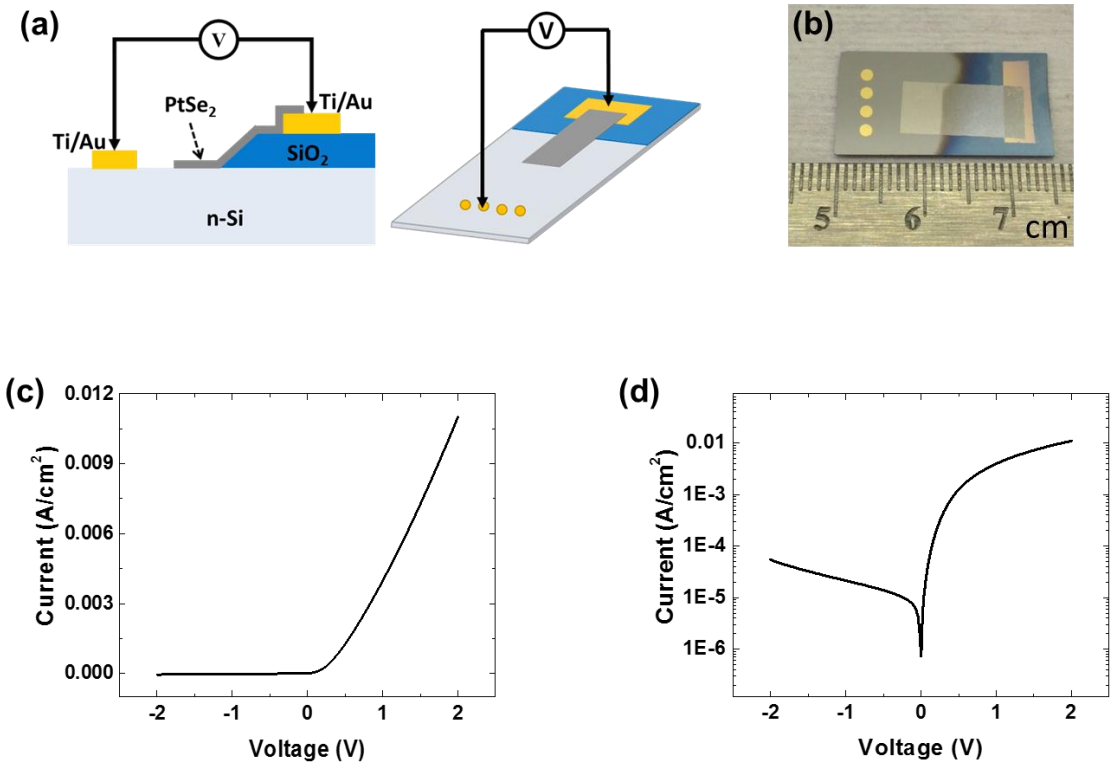

Figure 3. (a) Schematic diagram and (b) photograph of the PtSe$_2$/n-Si Schottky barrier diode. Current density-voltage (*J-V*) plots of the device with a PtSe$_2$ film synthesized from 4 nm thick Pt on a (c) linear and (d) semi-logarithmic scale.

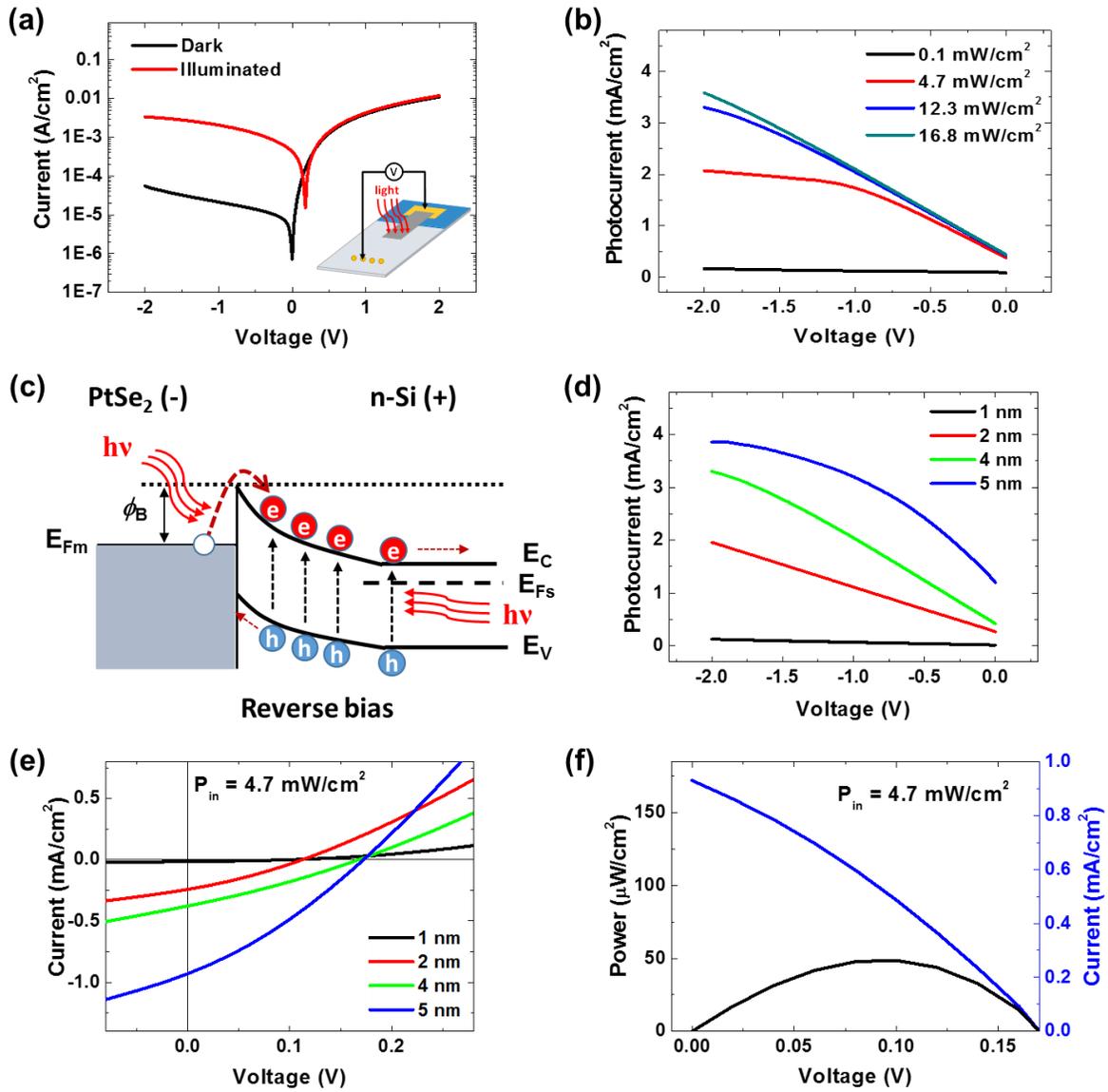

Figure 4. (a) *J-V* characteristics of the PtSe$_2$/n-Si SBD with a PtSe$_2$ film synthesized from 4 nm thick Pt in dark and under illumination and (b) its photocurrent plot under different light intensities in the reverse bias region. (c) Energy band diagram of the Schottky junction between PtSe$_2$ and n-Si under reverse bias. φ$_B$ and E$_{Fm}$/E$_{Fs}$ denote the Schottky barrier height and Fermi energy levels of PtSe$_2$/n-Si, respectively. E$_C$, E$_V$, and *hv* are the conduction band edge, valence band edge of n-Si, and the incident photon energy, respectively. (d) Photocurrent plot of the PtSe$_2$/n-Si SBDs with different PtSe$_2$ thicknesses (initial Pt deposition thicknesses are 1, 2, 4 and 5 nm, respectively) illuminated at the same incident light intensity (P$_{in}$ = 12.3 mW/cm$^2$). (e) *J-V* characteristics of the open circuit voltage (V$_{OC}$) and short circuit current (J$_{SC}$) for the PtSe$_2$/n-Si SBDs with different PtSe$_2$ thicknesses under an incident light intensity (P$_{in}$) of 4.7 mW/cm$^2$, and (f) plots of the maximum output power and corresponding *J-V* data of the PtSe$_2$/n-Si SBD with the PtSe$_2$ film synthesized from Pt of 5 nm starting thickness.

**TABLES**

Table 1. Summary of diode parameter values for the PtSe$_2$/n-Si SBDs, including the ideality factor ($n$), Schottky barrier height ($\varphi_B$), series resistance ($R_S$), open circuit voltage ($V_{OC}$), short circuit current ($J_{SC}$), maximum output power ($P_{max}$), fill factor (FF) and power conversion efficiency (PCE).

|  | PtSe$_2$/n-Si SBDs | | | |
| --- | --- | --- | --- | --- |
| **Pt Starting Thickness [nm]** | 1 | 2 | 4 | 5 |
| ***n*** | 1.7 | 1.4 | 1.9 | 1.6 |
| ***$\varphi_B$* [eV]** | 0.71 | 0.70 | 0.71 | 0.67 |
| ***$R_S$* [Ω]** | 1524 | 1121 | 836 | 453 |
| **$V_{OC}$ [V]** | 0.11 | 0.11 | 0.17 | 0.17 |
| **$J_{SC}$ [mA/cm$^2$]** | 0.02 | 0.24 | 0.38 | 0.93 |
| **$P_{max}$ [µW/cm$^2$]** | 0.52 | 8.16 | 18.03 | 48.66 |
| **FF** | 0.30 | 0.31 | 0.28 | 0.31 |
| **PCE [%]** | 0.01 | 0.17 | 0.38 | 1.04 |

## ASSOCIATED CONTENT

**Supporting Information**

The Supporting Information is available free of charge *via* the Internet at http://pubs.acs.org.

Additional details of the PtSe$_2$ film synthesis, selenization temperature study, current-voltage characteristics of the PtSe$_2$ gas sensors, limited switching speed of gas exchange and fitting model of the PtSe$_2$ sensors, and diode parameter extraction and photoconductivity of PtSe$_2$/n-Si Schottky barrier diodes, Figure S1 – S7 (PDF)

## AUTHOR INFORMATION

**Corresponding Authors**

*E-mail: nmcevoy@tcd.ie (N. McEvoy).

*E-mail: duesberg@tcd.ie (G. S. Duesberg).

**Author Contributions**

The manuscript was written through contributions of all authors. All authors have given approval to the final version of the manuscript.

#C.Y., K.L. and N.M. contributed equally to this work.

**Notes**

The authors declare no competing financial interest.

**ACKNOWLEDGMENTS**

This work is supported by the SFI under Contract No. 12/RC/2278 and PI_10/IN.1/I3030, and the European Union Seventh Framework Programme (Graphene Flagship, 604391). M.O.B. acknowledges an Irish Research Council scholarship *via* the Enterprise Partnership Scheme, Project 201517, Award 12508. N.M. acknowledges SFI (14/TIDA/2329). J.K. acknowledges funding from the Wiener Wissenschafts-, Forschungs-und Technologiefonds (WWTF) via project MA14-009. J.C.M. acknowledges support from the Austrian Science Fund (FWF) Project No. P25721-N20. M.C.L. acknowledges funding through an ERC grant (307311) and the German Research Foundation (DFG, LE 2440/1-2 and GRK 1564).

Supporting Information

# High-Performance Hybrid Electronic Devices from Layered PtSe$_2$ Films Grown at Low Temperature

*Chanyoung Yim,[†,‡,#] Kangho Lee,[†,#] Niall McEvoy,[\*,†,§,#] Maria O'Brien,[†,§] Sarah Riazimehr,[‡] Nina C. Berner,[†] Conor P. Cullen,[†,§] Jani Kotakoski,[Δ] Jannik C. Meyer,[Δ] Max C. Lemme,[‡] and Georg S. Duesberg[\*,†,§]*

[†]Centre for Research on Adaptive Nanostructures and Nanodevices (CRANN) and Advanced Materials and BioEngineering Research (AMBER), Trinity College Dublin, Dublin 2, Ireland

[‡]Department of Electrical Engineering and Computer Science, University of Siegen, Hölderlinstraße 3, 57076 Siegen, Germany

[§]School of Chemistry, Trinity College Dublin, Dublin 2, Ireland

[Δ]Faculty of Physics, University of Vienna, Boltzmanngasse 5, A-1090 Vienna, Austria

[#]C.Y., K.L. and N.M. contributed equally to this work.

[*]Corresponding authors: nmcevoy@tcd.ie and duesberg@tcd.ie.

**Film synthesis**

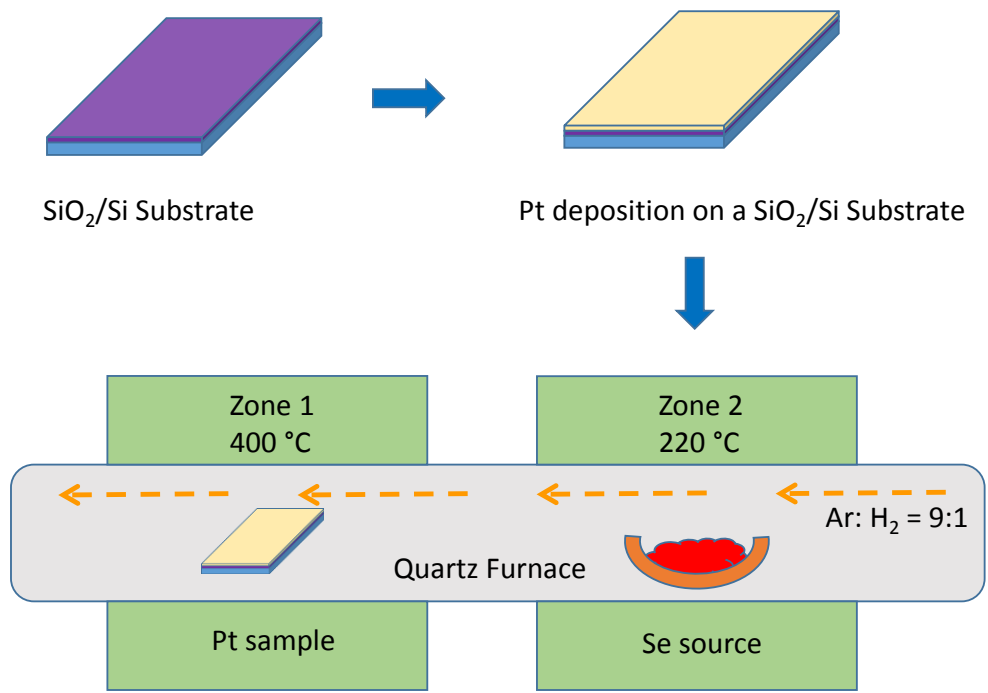

Figure S.1. Schematic diagram of the PtSe$_2$ film synthesis process using a vapor-phase selenization method.

**Selenization temperature study**

Pt samples were selenized at different temperatures in the range 250 – 400 °C keeping all other conditions described in the methods section identical. Raman spectra of films prepared in this way from starting Pt thicknesses of 0.5 nm and 3 nm are shown in Figure S.2 (a) and (b), respectively. In the case of the 0.5 nm films the spectra obtained for different selenization temperatures look similar, albeit with broader features seen for the 250 °C sample, suggesting lower crystallinity. However, in the case of the 3 nm films the spectra show notable differences with a clear increase in the relative intensity of the $A_{1g}$ mode seen with increasing selenization temperature. This trend is consistent with the formation of thicker PtSe$_2$ films at higher temperatures[1] implying that the Pt is not completely selenized at lower temperatures.

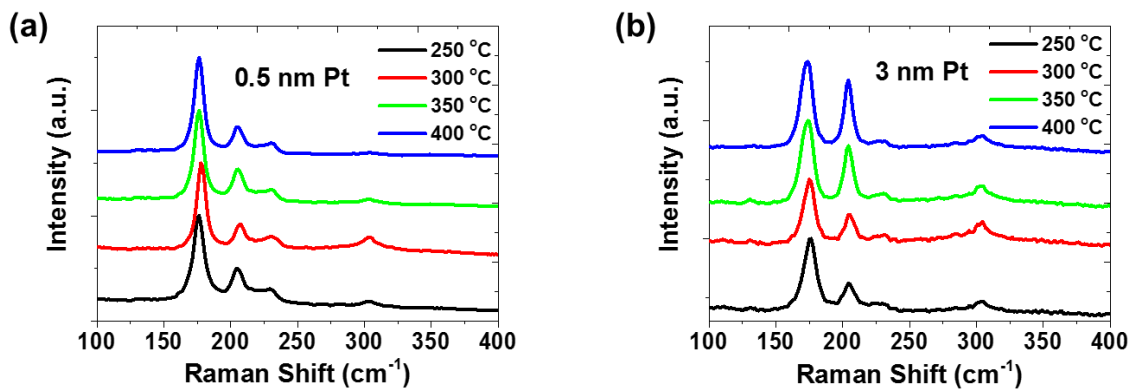

Figure S.2. Raman spectra of (a) 0.5 nm and (b) 3 nm Pt films selenized at different temperatures.

The 3 nm films were also characterized by X-ray photoelectron spectroscopy (XPS) as shown in Figure S.3. Fitting of the Pt 4f core level clearly shows strong contributions from unreacted Pt at selenization temperatures of 250 °C and 300 °C, suggesting incomplete selenization. For selenization temperatures of 350 °C and 400 °C the most prominent component observed is attributed to PtSe$_2$.

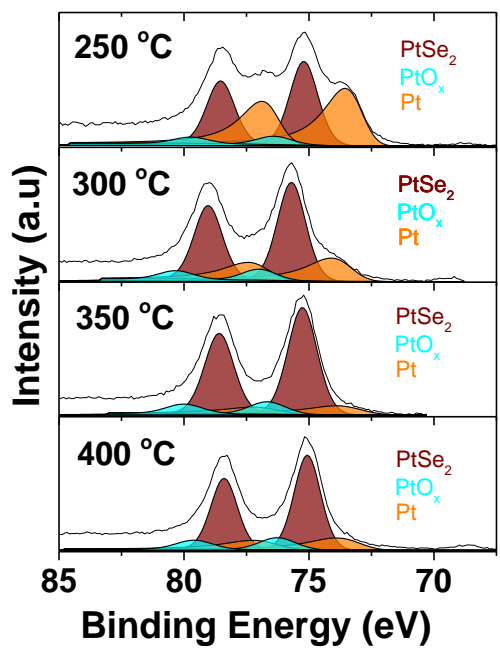

Figure S.3. XPS spectra showing the Pt 4f core level for 3 nm Pt films selenized at different temperatures.

**Current-voltage characteristics of the PtSe$_2$ sensors**

The contract resistance between the PtSe$_2$ channel and metal electrode (Ni/Au) of the gas sensor can be estimated by subtracting the channel resistance ($R_{Channel}$) from the total resistance ($R_{Total}$). The channel resistance was obtained by the four-terminal sensing method to eliminate contact resistance, and the total resistance was measured by the two-terminal sensing method. As shown in Figure S.4(a), our contacted devices have negligible contact resistance, approximately 184 kΩ·µm, which is only 2-3 % of channel resistance. In addition, the transfer characteristics of a transferred PtSe$_2$ film were examined at room temperature using a standard back-gate measurement configuration. Gate biases in the range of ± 100 V were applied to the PtSe$_2$ channel which exhibited p-type conduction, as depicted in Figure S.4(b).

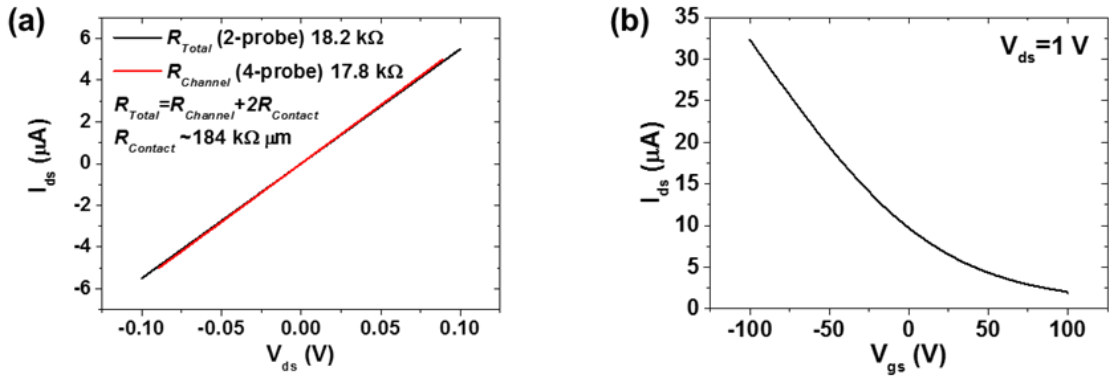

Figure S.4. (a) Plots of output characteristics in the configurations of 2- (black) and 4-probe measurements (red line). (b) A plot of transfer characteristics of a transferred PtSe$_2$ film in 2-probe measurement configuration.

**Limited switching speed of gas exchange and fitting model**

The PtSe$_2$ sensors immediately respond upon gas introduction. However, a multiplexer/switch, which was used for sequential measurements of multiple samples in our configuration, takes milliseconds for channel switching from sample to sample. Furthermore, mass-flow-controllers also take a few seconds to operate each proportional valve. As a result, the actual resistance change was observed a few seconds after gas introduction was set.

In comparison with a single exponential fitting, a double exponential model fits our data much better, as shown in Figure S.5

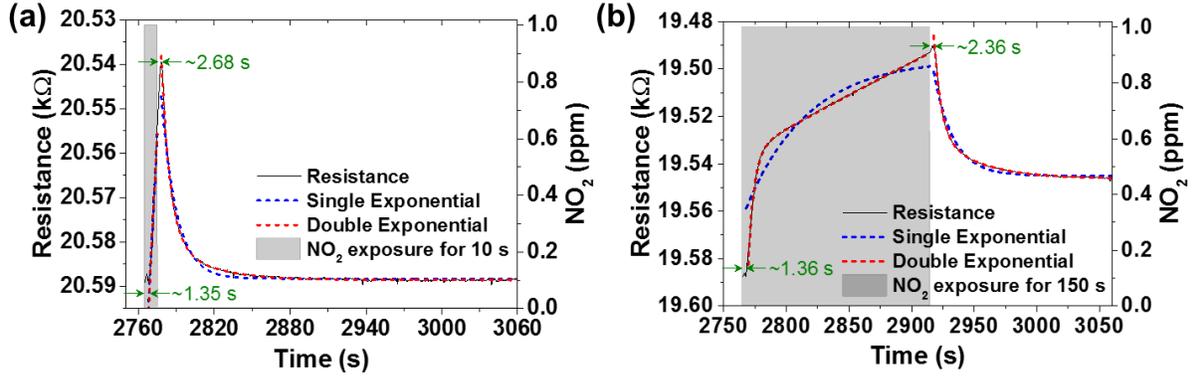

Figure S.5. Plots of resistance change upon (a) 10 s and (b) 150 s NO$_2$ gas exposure. Resistance change was monitored typically 1.35 and 2.50 seconds after gas introduction/cutout was set, due to the limited switching speed. The blue and red broken lines represent single and double exponential fitted curves, respectively.

**Diode parameter extraction and photoconductivity of PtSe$_2$/n-Si Schottky barrier diodes**

The current flow of the Schottky barrier diode (SBD) is based on thermionic emission theory. The current transport mechanism of the SBD can be explained by thermionic emission theory, which is given by the following equation,[2]

$$J = J_S[\exp\left(\frac{qV_D}{nk_BT}\right) - 1], \quad (S.1)$$

where $J$ is the output current density, $V_D$ is the voltage applied across the junction, $J_S$ is the reverse saturation current density, $n$ is the ideality factor, $k_B$ is the Boltzmann constant, $q$ is the elementary charge and $T$ is the absolute temperature in Kelvin. $J_S$ is expressed as

$$J_S = A^{**}T^2\exp(\frac{-q\varphi_B}{k_BT}), \quad (S.2)$$

where $A^{**}$ is the effective Richardson constant for n-Si (110 A cm$^{-2}$K$^{-2}$) and $\varphi_B$ is the effective Schottky barrier height (SBH) at zero bias. In a real diode, there is an undesirable resistance factor, called series resistance ($R_S$), which can cause degradation of the device performance. It is almost impossible to completely rule out the effect of $R_S$ in a real device operation. When considering the effect of $R_S$ of the system, $V_D$ of the Eq. (S.1) can be expressed by the combination of the total voltage drop of the system ($V$) and the voltage drop induced by $R_S$ ($JR_S$). Therefore, for $V_D > 3k_BT/q$, Eq. (1) becomes

$$J = J_S[\exp\left(\frac{q(V-JR_S)}{nk_BT}\right)]. \tag{S.3}$$

Using Cheung's method,[3] Eq. (S.3) can be rewritten as

$$V = JAR_S + n\varphi_B + \left(\frac{nk_BT}{q}\right)\ln\left(\frac{J}{A^{**}T^2}\right), \tag{S.4}$$

where $A$ is the effective diode area.

Differentiating Eq. (S.4) with respect to the current density $J$, it can be rewritten as

$$\frac{dV}{d(\ln J)} = JAR_S + \frac{nk_BT}{q}. \tag{S.5}$$

Using the linearity of the plot of $dV/d(\ln J)$ vs. $J$ in Eq. (S.5), $R_S$ and $n$ can be extracted from the slope and the y-axis intercept, respectively.

In addition, the auxiliary equation $H(J)$ can be defined from Eq. (S.4) as

$$H(J) = V - \left(\frac{nk_BT}{q}\right)\ln\left(\frac{J}{A^{**}T^2}\right), \tag{S.6}$$

$$H(J) = JAR_S + n\varphi_B. \tag{S.7}$$

Using the linearity of the plot of *H(J)* vs. *J* from Eq. (S.7), $\varphi_B$ can be found from the y-axis intercept of the plot.

The representative plots of *dV/d(lnJ)* vs. *J* and *H(J)* vs. *J* for the SBD with a PtSe$_2$ film synthesized by the selenization of a 4 nm thick Pt film are presented in Figure S.6.

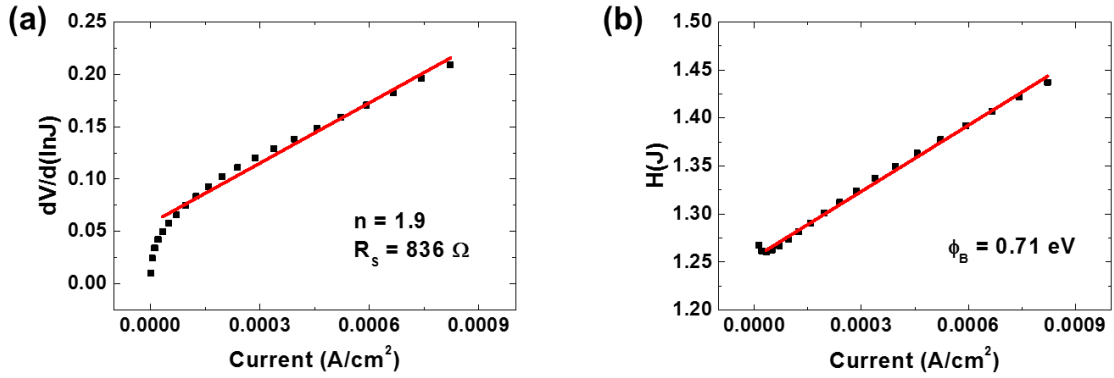

Figure S.6. Plots of (a) *dV/d(lnJ)* vs. *J* and (b) *H(J)* vs. *J* for the SBD with a PtSe$_2$ film synthesized from 4 nm thick Pt, giving the values of the ideality factor, series resistance and the Schottky barrier height of the device.

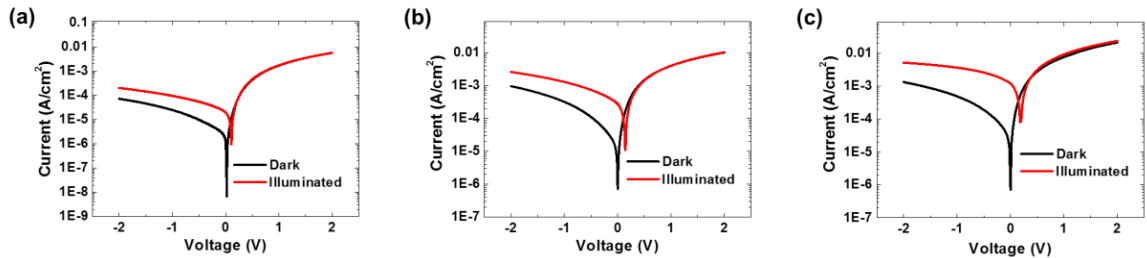

Figure S.7. Plots of current-voltage data measured under the dark and illuminated conditions ($P_{in}$ = 12.3 mW/cm$^2$) for the PtSe$_2$/n-Si SBDs with PtSe$_2$ films synthesized from (a) 1 nm, (b) 2 nm and (c) 5 nm thick Pt.